\documentclass[twocolumn,amsmath,amssymb,prl,superscriptaddress,floatfix,showpacs]{revtex4}
\usepackage{graphicx}
\usepackage{color}
\usepackage{hyperref}

\newcommand{\vlowk}{\ensuremath{V_{\text{low }k}}}

\graphicspath{{./}{./figures/}{./figures/iso3/}{./figures/iso1/}}

\begin{document}
\title{Low-momentum Hyperon-Nucleon Interactions}

\author{B.-J. Schaefer}
\email[E-Mail:]{bernd-jochen.schaefer@physik.tu-darmstadt.de}
\author{M. Wagner}
\affiliation{Institut f\"{u}r Kernphysik, TU Darmstadt, D-64289 Darmstadt, Germany}
\author{J. Wambach}
\affiliation{Institut f\"{u}r Kernphysik, TU Darmstadt, D-64289 Darmstadt, Germany}
\affiliation{Gesellschaft f\"{u}r Schwerionenforschung GSI, D-64291 Darmstadt, Germany}
\author{T.T.S. Kuo}
\author{G.E. Brown}
\affiliation{Department of Physics and Astronomy, State University of New York, Stony Brook, NY 11794-3800, USA}

\date{\today}


\begin{abstract}
  We present a first exploratory study for hyperon-nucleon
  interactions using renormalization group techniques. The effective
  two-body low-momentum potential $\vlowk$ is obtained by integrating
  out the high-momentum components from realistic Nijmegen $YN$
  potentials. A $T$-matrix equivalence approach is employed, so that
  the low-energy phase shifts are reproduced by $\vlowk$ up to a
  momentum scale $\Lambda \sim 500$ MeV. Although the various bare
  Nijmegen models differ somewhat from each other, the corresponding
  $\vlowk$ interactions show convergence in some channels, suggesting
  a possible unique $YN$ interaction at low momenta.
\end{abstract}

\maketitle
\newpage

Starting from modern nucleon-nucleon interactions and performing a
renormalization group (RG) decimation it has become possible to derive
a unique low-momentum effective interaction $\vlowk$~\cite{SKB03}. The
basic idea is to integrate out the short-distance physics encoded in
hard-core interactions which are not well constrained by the available
phase shift data. The resulting effective interactions form the
starting point for ab-initio nuclear structure calculations in
few-body systems~\cite{ANSKB04}, shell-model
studies~\cite{Andreozzi:1996tf} and mean-field treatments via
density-functional methods~\cite{Bogner:2005sn}.  They also serve as
input for the derivation of Landau Fermi-liquid interactions and
provide predictions of pairing gaps in nuclei and homogeneous neutron
matter~\cite{Schwenk2002}.

In this paper we generalize the $\vlowk$ approach to the
hyperon-nucleon ($YN$) sector. The ultimate goal is to provide
effective potentials of similar quality as in the $NN$ case that could
serve as the starting point for realistic calculations of the
structure of hypernuclei and homogeneous hyperonic matter. At present
such a program is hampered by the lack of a comparable data base and
the collapse to a unique low-momentum potential is far from
obvious. In a first exploratory study we wish to address this point by
considering the low-momentum decimation of various potentials by the
Nijmegen group. We focus on hyperons with strangeness $S=-1$ for which
$I=1/2$ and $I=3/2$ isospin states are available. For $I=1/2$, several
hyperon-nucleon channels occur which require new technical
developments for the coupled RG flow equations.
\vspace{1ex}

The effective, low-momentum potential $\vlowk$ for elastic two-body
scattering is obtained by integrating out high-momentum components of
a realistic bare potential $V$ interaction. This is achieved by
imposing a cutoff $\Lambda$ on all loop integrals in the half-on-shell
(HOS) $T$-matrix equation and replacing the bare potential $V$ with
the effective $\vlowk$ potential. Since the physical low-energy
quantities must not depend on the cutoff, the HOS $T$-matrix should be
preserved for relative three-momenta $k'$,$k \leq \Lambda$. This
results in a modified Lippmann-Schwinger equation with a
cutoff-dependent effective potential $\vlowk$
\[
  T(k'\!\!,k;k^2)\!=\!\vlowk(k'\!\!,k)+\frac 2 \pi{\cal
P}\!\!\!\int\limits_0^{\Lambda}\!\! q^2 dq \frac{\vlowk (k'\!\!,q) T
(q,\!k;k^2)} {k^2-q^2}.
\]

By demanding $dT(k',k;k^2)/d\Lambda = 0$, an exact renormalization
group (RG) flow equation for $\vlowk$ can be obtained
\cite{SKB+01}
\begin{equation}
\label{rgeq}
\frac {d}{d \Lambda} \vlowk (k',k) = \frac 2 \pi \frac{\vlowk (k',
  \Lambda) T(\Lambda,k;\Lambda^2)}{1-k^2/\Lambda^2}\ .
\end{equation}

Integrating this flow equation with a given initial bare potential at
a large cutoff (small distance) one obtains the physically equivalent
effective theory ($\vlowk$) at a smaller cutoff $\Lambda$ (larger
distance).

Instead of solving the RG equation (\ref{rgeq}) directly as a
differential equation with e.g. standard Runge-Kutta methods, we use
the Andreozzi-Lee-Suzuki (ALS) iteration method, which is based on a
similarity transformation \cite{andreozzi,suzuki80}. With folded
diagram techniques it has been shown \cite{bogner02} that this
iteration method indeed yields a solution of Eq.~(\ref{rgeq}).

By construction, the resulting $\vlowk$ for the $NN$ interaction
reproduces the empirical deuteron binding energy and scattering phase
shifts up to $E_{lab}= 2\hbar^2\Lambda^2/M$. Most importantly, it is
found that for $\Lambda < 2$ fm$^{-1}$ ($E_{lab}< 330$ MeV) $\vlowk$
is independent of the particular $V_{NN}$ model, i.e. all diagonal
matrix elements of the different high-precision potentials collapse to
a single unique low-momentum effective potential~\cite{SKB+01}. This
can be largely attributed to the long-range one-pion exchange (OPE)
which is common to all realistic potentials and dominates the
low-momentum scattering. The main effect of the RG evolution is a
constant shift of the bare matrix elements which removes the
ambiguities in the short-range part of the potential.

The energy-independent $\vlowk$ is non-Hermitian. This can readily be
seen from the RG equation~(\ref{rgeq}) because the momenta are treated
asymmetrically. With a second similarity transformation the
non-Hermiticity of $\vlowk$ can be eliminated.  Phase shifts are
preserved by this second transformation and there are a number of such
phase shift equivalent transformations such as the well-known Okubo
one~\cite{holt04}. In the present work we have used the Okubo
transformation to obtain the Hermitian $\vlowk$. For the $NN$
interaction, the diagonal matrix elements are almost unchanged by the
second transformation.


As realistic $YN$ interactions we use the soft-core potentials by the
Nijmegen group~\cite{rijken}. They are based on one-boson-exchange
models (OBE) of the $NN$ potential and use $SU(3)_F$-symmetry to infer
the coupling vertices in the presence of a hyperon. Since the flavor
symmetry is broken by the finite quark masses the pertinent coupling
strengths have to be adjusted to data. Six different fits are
available referred to NSC97a - NSC97f in the following.  Each
potential comes in two basis representations (isospin- and physical
particle-basis). In this paper we work on the isospin basis which was
also originally used for the potential construction by the Nijmegen
group. Therefore all isomultiplets are degenerate. The corresponding
isospin-averaged masses are given by $m_N = 938.9$ MeV, $m_\Lambda =
1115.7$ MeV and $m_\Sigma = 1193.1$ MeV. In the $I=3/2$ channel no
$\Lambda$-hyperon is involved.

The different Nijmegen fits describe the known $YN$ cross section data
equally well ($\chi^2/N \sim 0.55$) but exhibit differences on a more
detailed level. Due to the few available data points (only 35
altogether), the phase shifts and some scattering lengths exhibit
large variations for different fits~\cite{nnon}. This is in marked
contrast to the $NN$ case where the wealth of the data base allows for
high-precision potentials. These differ only in their short-range
properties which are not constrained by the available data.  The
collapse of the low-momentum potential $\vlowk$ for the $NN$
interaction to a 'unique' effective potential, observed after the RG
decimation~\cite{SKB03} is basically driven by the precision of the
measured phase shifts. Due to the scarceness and uncertainties in the
$YN$ data we cannot expect a unique low-momentum $YN$ $\vlowk$
potential.

In addition, for the $YN$ interaction the choice of the cutoff
$\Lambda $ is not so obvious. In this work a cutoff around $\Lambda
\sim 500$ MeV is chosen which corresponds to $2.5$ fm$^{-1}$ in the
configuration space.
\begin{figure}[!htb]
\centerline{\hbox{
  \includegraphics[width=0.5\columnwidth]{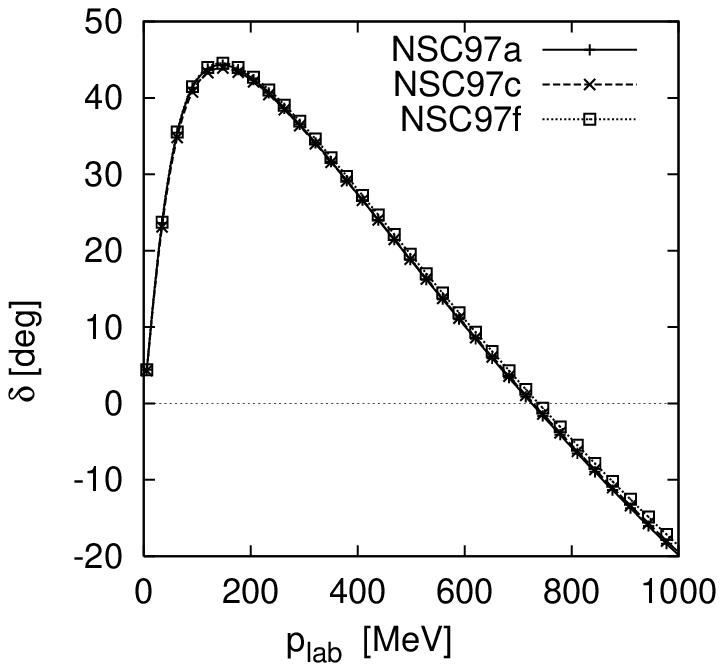}
  \includegraphics[width=0.5\columnwidth]{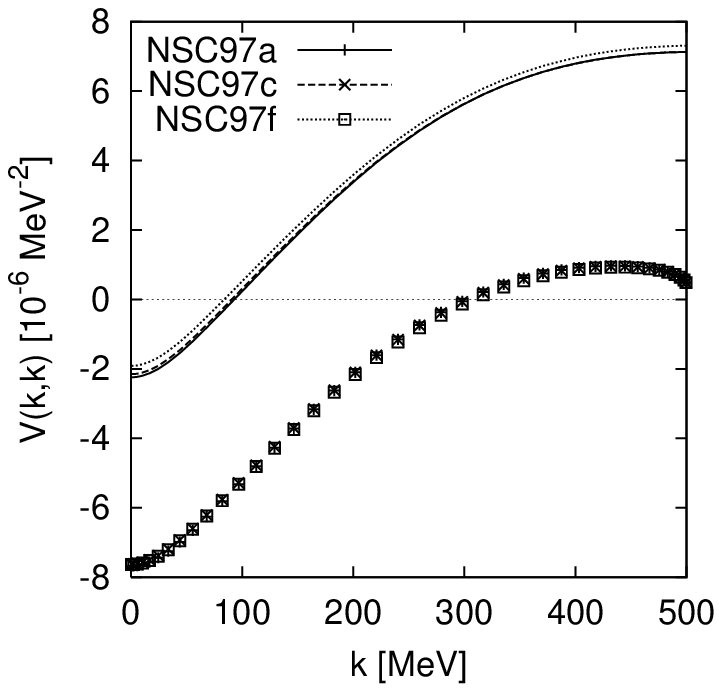}
  }}
\caption{\label{fig:sig1s0} Left panel: The corresponding $^1S_0$
phase shifts for the six different potentials as a function of the
momentum in the lab frame $p_{\rm LAB}$. Right panel: Three diagonal
bare potentials $V_{\rm bare}$ (NSC97a,c,f; dashed lines) and three
(Hermitian) $\vlowk$ matrix elements (dotted) for $\Sigma N \to \Sigma
N$ ($I = 3/2$, $^1S_0$) versus the relative momentum $k$. }
\end{figure}
\vspace{1ex}

We first consider the $I=3/2$ case which corresponds to the $\Sigma N
\to \Sigma N$ channel. The ALS iteration is exactly the same as for
the $NN$ interaction albeit with the appropriate substitution of the
hyperon mass. For the ALS iteration we have used a cutoff $\Lambda
\equiv\Lambda_P = 500$ MeV in the model space ($P$-space) with $64$
grid points and a cutoff $\Lambda_Q = 10$ GeV for the complementary
$Q$-space also with $64$ grid points. In order to verify the
insensitivity of the results on these quantities we have varied the
grid points in an interval $[32,70]$ and the cutoff $\Lambda_Q$ in
between the range $\Lambda_Q\pm 2$ GeV. The standard ALS method
converges rapidly. Details will be presented in~\cite{2letter}.

The results for the $^1S_0$ partial wave are shown in
Fig.~\ref{fig:sig1s0}.  In the right panel the diagonal matrix
elements for three different bare potentials (dashed lines) and the
corresponding RG evolved $\vlowk$ potentials are displayed versus the
relative momentum $k$. The RG decimation yields a soft-core $\vlowk$
potential which is more attractive (dotted curves). As already
mentioned, the $\vlowk$'s are basically non-Hermitian after the RG
decimation. By means of an Okubo-transformation we obtain Hermitian
$\vlowk$ potentials. As in the $NN$ case~\cite{SKB03} the differences
are negligible.

By construction, the on-shell $T$-matrix is phase-shift equivalent
which must result in identical phase shifts for a given bare potential
fit and momenta below the cutoff $\Lambda$.  This is demonstrated in
the left panel of Fig.~\ref{fig:sig1s0}. This comparison also serves
as a test for our numerics. For the $^1S_0$ partial wave, the bare
potentials do not differ strongly and thus yield almost the same
$\vlowk$ for all fits considered.
\begin{figure}[!htb]
  \centerline{\hbox{
  \includegraphics[width=0.5\columnwidth]{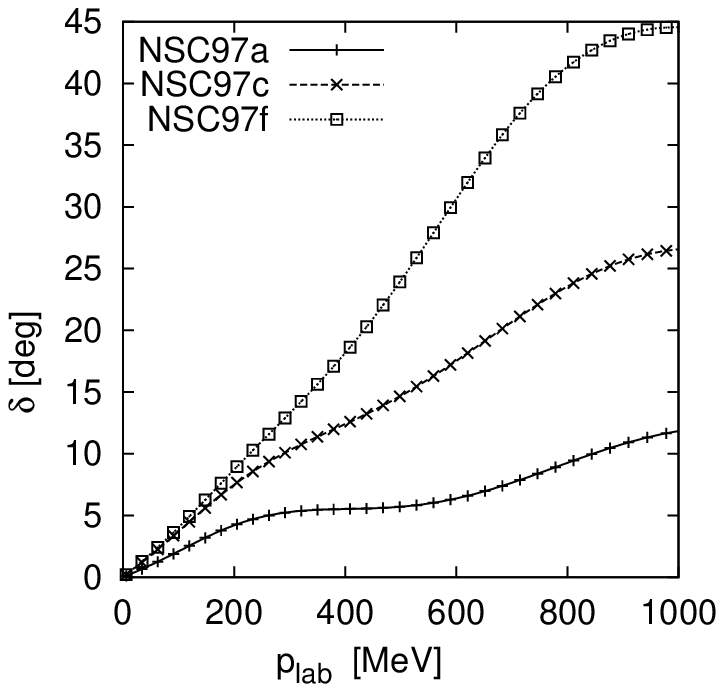}
  \includegraphics[width=0.5\columnwidth]{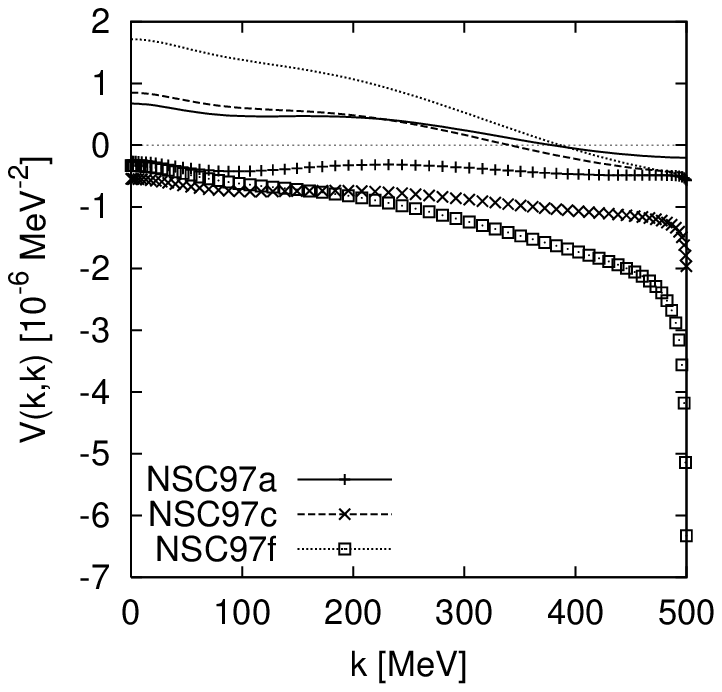}
  }}
\caption{\label{fig:sigtensor} The same as in Fig.~\ref{fig:sig1s0}
for the partial wave $^3S_1$. Left panel: nuclear bar phase shift.}
\end{figure}
As discussed below, this changes for higher partial waves where the
different bare potential fits deviate significantly and therefore no
'unique' $\vlowk$ is found.

As in the $NN$ case, the $YN$ interaction contains tensor components
and hence partial waves can mix. The generalization of the ALS
iteration in the presence of tensor forces is straightforward. For
$S=1$ one has to enlarge the $T$-matrix to a $(2\times 2)$ block
structure in the standard way, corresponding to the orbital angular
momentum combinations $L=J\pm 1$. Also for this case we have verified
that $\vlowk$ is phase-shift equivalent to the bare potentials, as is
shown in left panel Fig.~\ref{fig:sigtensor} for the $^3S_1$ partial
wave.  The right panel displays the corresponding $\vlowk$ potentials
together with the bare ones.  They again become more attractive for
all Nijmegen fits. Due to strong differences in the phase shifts we do
not find a collapse of $\vlowk$ to one potential, especially at larger
momenta where deviations are most pronounced.

The RG flow equation (\ref{rgeq}) implies a pole at the cutoff
boundary $k=\Lambda$. In the vicinity of this pole the slope for the
$\vlowk$ diverges which can be clearly seen in the right panel of
Fig.~\ref{fig:sigtensor}.
\begin{figure}[!htb]
  \centerline{\hbox{
  \includegraphics[width=0.5\columnwidth]{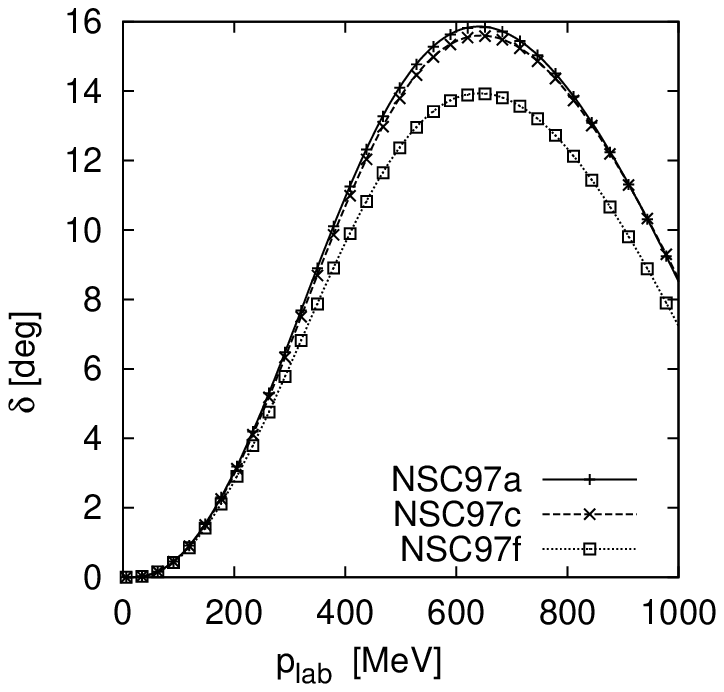}
  \includegraphics[width=0.5\columnwidth]{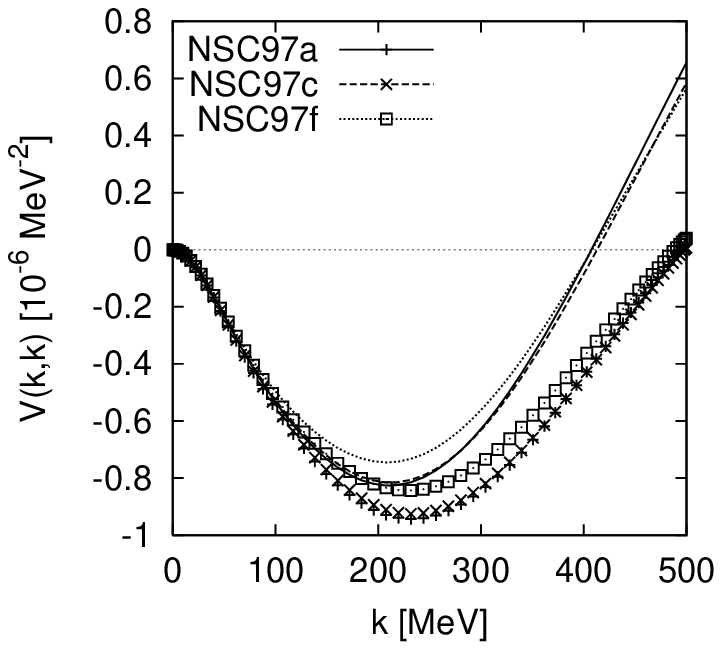}
        }}\caption{\label{fig:sig1p1} Same as
        Fig.~\ref{fig:sig1s0} for the partial wave $^1P_1$ and
        $I=3/2$.}
\end{figure}

To complete the analysis for the $I=3/2$ channel we show in
Fig.~\ref{fig:sig1p1} the $^1P_1$ partial wave which serves as an
example for the spin singlet-spin triplet transition. Such transitions
which are induced by the tensor force are negligible in the $NN$
interaction due to the small mass differences. However, for the $YN$
interaction this is not the case anymore and these transitions can be
significant. Due to the smaller deviations in the phase shift, the
corresponding $\vlowk$ interactions collapse to single potential
(especially for small momenta).
\vspace{1ex}

The treatment of of the $I=1/2$ channel is much more complicated.  For
this isospin four coupled channels available corresponding to the
transitions: $(\Lambda N,\Sigma N )\to(\Lambda N,\Sigma N )$. This is
a completely new situation for the $\vlowk$ approach. Since now
channels with different masses couple, new phenomena are to be
expected concerning e.g.~the convergence behavior of the ALS
iteration.

In flavor space the Lippmann-Schwinger equation becomes a coupled
($2\times 2$) matrix equation where the diagonal matrix elements
describe respectively the $\Lambda N \to\Lambda N$ and $\Sigma N \to
\Sigma N$ channels while the off-diagonal elements describe the
$\Lambda N \to \Sigma N$ and $\Sigma N \to\Lambda N$ transitions.
Using a notation where we only list the hyperons ($Y,Y'=\Lambda,
\Sigma$)
\begin{eqnarray}
\label{LS}
T^{Y'Y}(k',k;E_k^Y) = V^{Y'Y}(k',k) \hspace*{3cm}&& \nonumber \\
+\sum_{Z=\Lambda,\Sigma}
\frac{2}{\pi} \mathcal{P}\!\! \int\limits_0^\infty\!\! dq q^2
\frac{V^{Y'Z}(k',q) T^{ZY}(q,k;E_k^Y)}
{E^Y_k -H^Z_0 }&&
\end{eqnarray}
with the free Hamiltonian $H^Y_0(q)=\frac {q^2}{2\mu_Y} + m_Y + m_N$,
the energy $E_k^Y = \frac {k^2}{2\mu_Y} + m_Y + m_N$ and the reduced
mass $\mu_Y = \frac {m_Y m_N}{m_Y+m_N}$ we have (with the inclusion of
tensor forces) in general four coupled equations to solve. For some
channels, the mass difference $m_\Sigma - m_\Lambda$ enters in the
denominator of Eq.~(\ref{LS}) which induces e.g.~a threshold behavior
for the $\Sigma$ hyperon. The mass differences which are not present
in the $NN$ interaction enter also in the ALS iteration. It is found
that the standard ALS iteration procedure does not converge to the
proper consecutive set of eigenvalues. As a consequence, a wrong
sorting of the eigenvalues in the different $P$- and $Q$-spaces
emerges. Using a modified ALS iteration or by introducing an energy
cutoff solves this problem and convergence to the correct eigenvalues
can be found~\cite{2letter}.

\begin{figure}[!htb]
  \centerline{\hbox{
  \includegraphics[width=0.5\columnwidth]{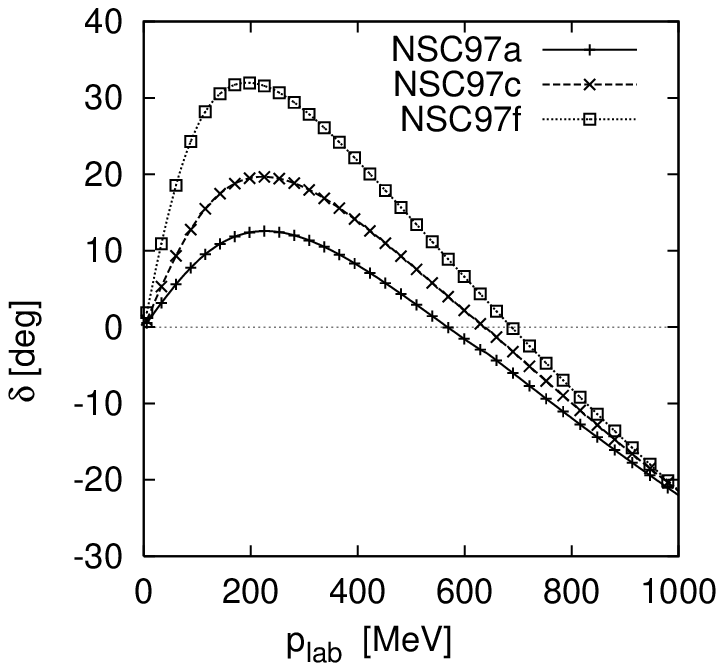}
  \includegraphics[width=0.5\columnwidth]{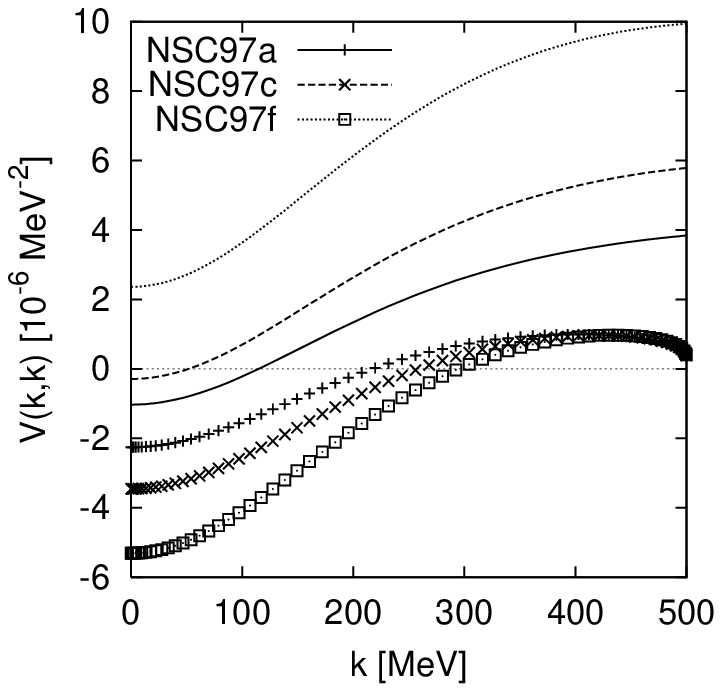} }}
  \caption{\label{fig:lam1s0} $^1S_0$ partial wave for the $I=1/2$
  $\Lambda N \to \Lambda N$ channel. Left panel: Corresponding phase
  shifts.  Right panel: bare potentials and $\vlowk$ potentials. The
  labeling in both panels is the same as in Fig.~\ref{fig:sig1s0}.}
\end{figure}

As an example we show in Fig.~\ref{fig:lam1s0} the $^1S_0$ partial
wave for the $\Lambda N \to \Lambda N$ channel. All $\vlowk$
potentials are again more attractive and a more narrow grouping as
compared to the bare potentials can be observed.  One also observes
that the shift of the $\vlowk$ potential is largest for the bare
NSC97f potential and smallest for the NSC97a potential in contrast to
all other partial waves. The $\vlowk$'s are shifted in such a way that
the potentials collapse for relative momenta near the cutoff
reflecting the corresponding trends in the phase shifts.
\begin{figure}[!htb]
\centerline{\hbox{
\includegraphics[width=0.5\columnwidth]{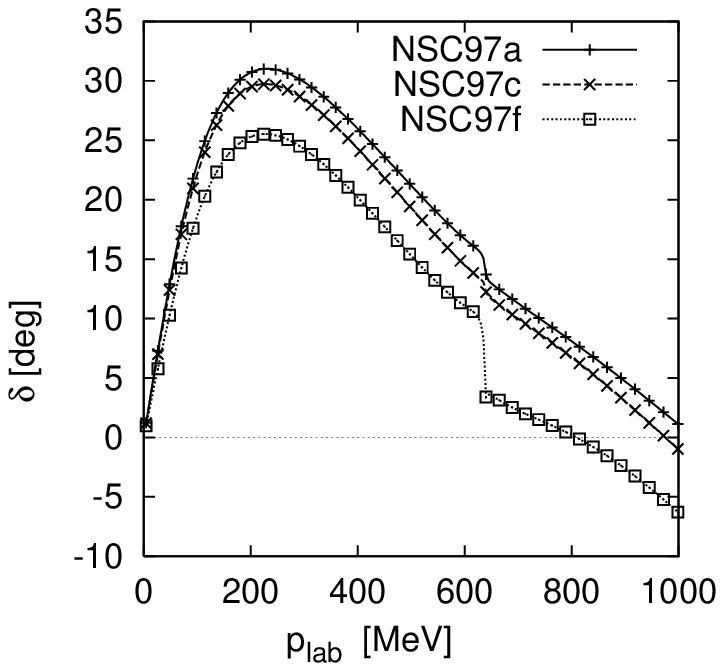}
\includegraphics[width=0.5\columnwidth]{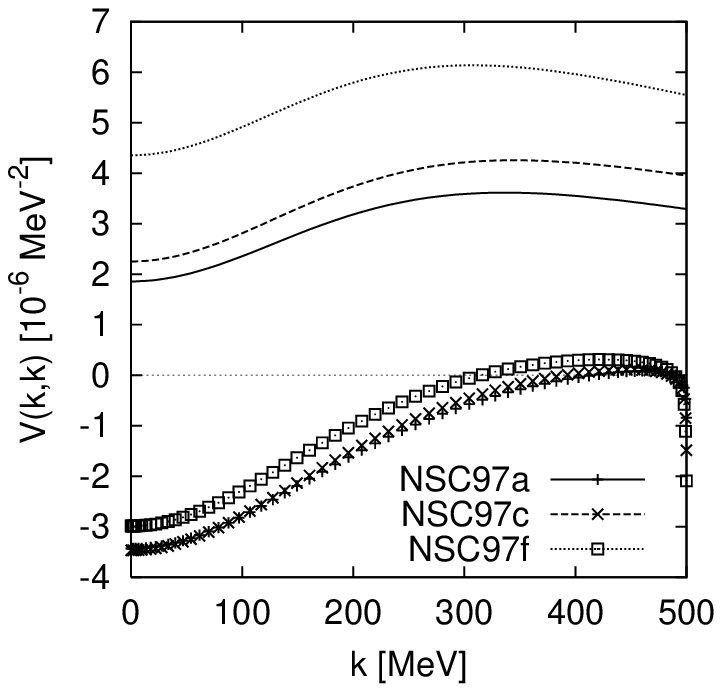}
}} \caption{\label{fig:lamtensor} Same as Fig.~\ref{fig:lam1s0} but
for the partial wave $^3S_1$. Left panel: nuclear bar phase shifts.}
\end{figure}

In Fig.~\ref{fig:lamtensor} the $\Lambda N \to \Lambda N$ channel with
tensor coupling is shown for the $^3S_1$ partial wave . Here again the
RG decimation pushes the bare potentials down basically by a (large)
constant to attractive $\vlowk$'s. Since this channel includes the
$\Sigma N$ transition, the $\Sigma$ threshold is visible in the phase
shifts for lab-momenta above $600$ MeV. The jump in the phase shift at
the threshold depends strongly on the model.
\begin{figure}[!htb]
\centerline{\hbox{
\includegraphics[width=0.5\columnwidth]{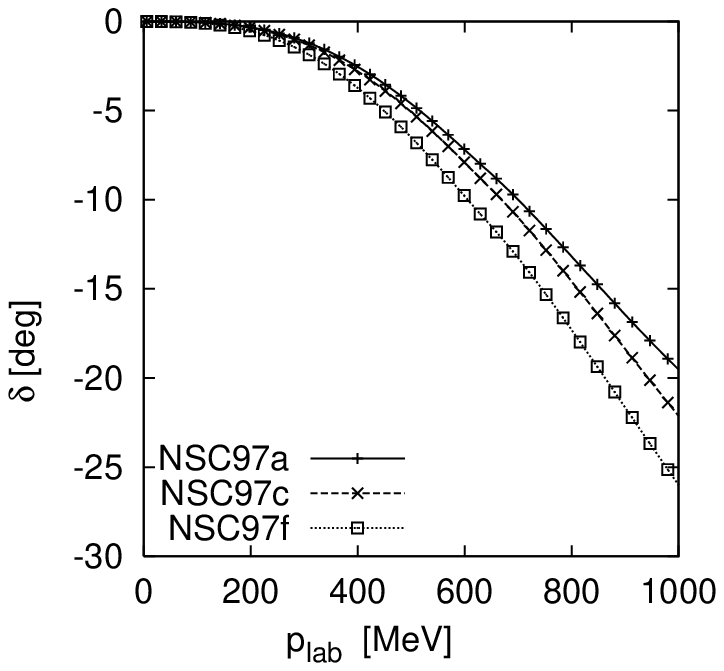}
\includegraphics[width=0.5\columnwidth]{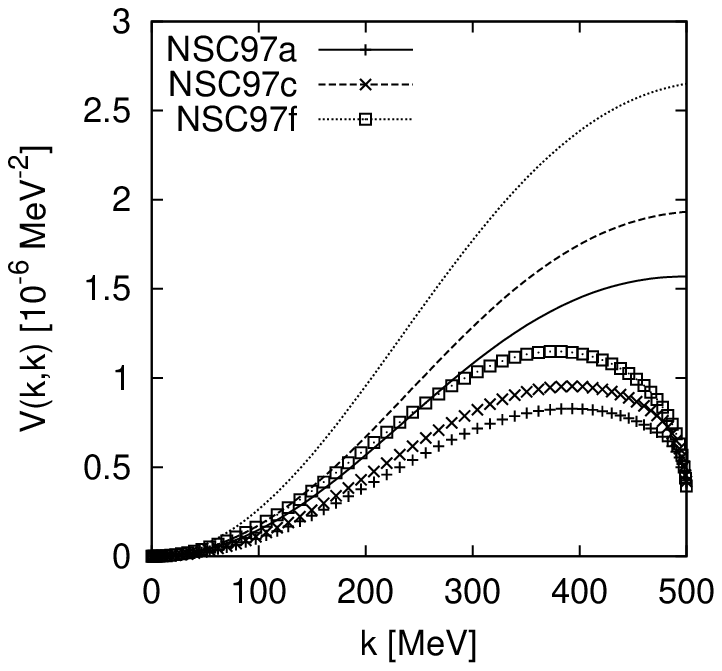}
}}\caption{\label{fig:lam1p1} Same as Fig.~\ref{fig:lam1s0} but for
the partial wave $^1P_1$. Left panel: nuclear bar phase shift.}
\end{figure}

To complete this analysis results for the spin singlet - spin triplet
transition in the $\Lambda N \to \Lambda N$ channel for the $^1P_1$
partial wave are presented in Fig.~\ref{fig:lam1p1}. For small momenta
no RG decimation takes place. For this partial wave the bare
interaction is repulsive and so is $\vlowk$.

\vspace{1ex}

Recently, it has been shown that the $\vlowk$ approach which is based
on RG techniques provides a novel and powerful tool to obtain
phase-shift equivalent low-momentum $NN$ interactions. After the RG
decimation a unique low-momentum potential $\vlowk$ for different
high-precision $NN$ interactions was found. This model-independence of
the diagonal matrix elements of the $\vlowk$ is an important property
which is basically driven by the phase shift equivalence of the input
models.

In the present work the model-dependence of the $\vlowk$ for the $YN$
interaction is investigated. Because of coupled-channel effects in
flavor space that are not present in the $NN$ case, the RG evolution
is technically more complicated but can be treated. Due to the few
experimental data currently available, the model fits by the Nijmegen
group do not allow for a unique low-momentum $YN$ interaction. It is
therefore of importance to calculate $\vlowk$ for other $YN$ models
such as the revised J\"ulich potential~\cite{JHWM01}. Such
calculations are in preparation~\cite{2letter}.

Although no unique $YN$ low-momentum interaction is obtained at
present, a convergence of the different $\vlowk$'s is seen generally
for all the Nijmegen potentials. Especially, for partial waves which
do not deviate strongly for different bare potentials the uniqueness
of $\vlowk$ is pronounced. All $\vlowk$ potentials are much softer
than the bare ones. Softer interactions lead to stronger binding which
should be of relevance in microscopic hypernuclei calculations.

By construction, all low-energy two-body observables are
cutoff-independent. Bogner et al. argue that any (new) induced cutoff
dependence is due to higher-body forces~\cite{ANSKB04}. For the $NN$
interaction they conclude that such contributions are rather
small. For the $YN$ interaction this is still an open issue and should
be tested in light hypernuclei.

We thank A. Schwenk for helpful discussions. One of the authors (BJS)
would also like to thank S.K.~Bogner for numerous enlightening
discussions. He also expresses his gratitude to G.E.~Brown and
T.T.S.~Kuo for the invitation to Stony Brook where this work was
initiated. MW is supported by BMBF Grand No.~06DA116.  Partial support
to TTSK and GEB from the US Department of Energy under contract
DE-FG02-88ER/40388 is gratefully acknowledged.



\end{document}